# Surface Dyakonov-Cherenkov Radiation

Hao Hu[1,#], Xiao Lin[2,3,#], Liang Jie Wong[1], Qianru Yang[1], Baile Zhang[4,5,*], and Yu Luo[1,*]

[1]School of Electrical and Electronic Engineering, Nanyang Technological University, Nanyang Avenue, Singapore 639798, Singapore.

[2]Interdisciplinary Center for Quantum Information, State Key Laboratory of Modern Optical Instrumentation, ZJU-Hangzhou Global Science and Technology Innovation Center, College of Information Science and Electronic Engineering, Zhejiang University, Hangzhou 310027, China.

[3]International Joint Innovation Center, ZJU-UIUC Institute, Zhejiang University, Haining 314400, China.

[4]Division of Physics and Applied Physics, School of Physical and Mathematical Sciences, Nanyang Technological University, 21 Nanyang Link, Singapore 637371, Singapore.

[5]Centre for Disruptive Photonic Technologies, Nanyang Technological University, Singapore 637371, Singapore.

[#]These authors contributed equally.

[*]E-mail: blzhang@ntu.edu.sg (B. Zhang); luoyu@ntu.edu.sg (Y. Luo)

**Recent advances in engineered material technologies (e.g., photonic crystals, metamaterials, plasmonics, etc) provide valuable tools to control Cherenkov radiation. In all these approaches, however, the designed materials interact only with the particle velocity to affect Cherenkov radiation, while the influence of the particle trajectory is generally negligible. Here, we report on surface Dyakonov-Cherenkov radiation, i.e. the emission of directional Dyakonov surface waves from a swift charged particle moving atop a birefringent crystal. This new type of Cherenkov radiation is highly susceptible to both the particle velocity and trajectory, e.g. we observe a sharp radiation enhancement when the particle trajectory falls in the vicinity of a particular direction. Moreover, close to the Cherenkov threshold, such a radiation enhancement can be orders of magnitude higher than that obtained in traditional Cherenkov detectors. These distinct properties allow us to determine simultaneously the magnitude and direction of particle velocities on a compact platform. The surface Dyakonov-Cherenkov radiation studied in this work not only adds a new degree of freedom for particle identification, but also provides an all-dielectric route to construct compact Cherenkov detectors with enhanced sensitivity.**

Cherenkov radiation arises when a charged particle moves with a velocity exceeding the phase velocity of light in a transparent medium [1,2]. In transparent dielectric materials, Cherenkov radiation is an extended mode (i.e., a packet of Cherenkov photons) propagating into the infinity at a constant emission angle $\theta_c$. This so-called Cherenkov angle depends on the particle velocity $v$ as $\cos\theta_c = \frac{c}{nv}$ [3], where $c$ is the light speed in the vacuum and $n$ is the refractive index of the background medium. As a key application in particle physics, Cherenkov radiation has been widely deployed for particle discrimination, i.e. the particle velocity can be determined with high accuracy by measuring the Cherenkov angle of emitted photons [4-10]. Emerging communication applications call for a route map towards the miniaturization of Cherenkov devices [11-15]. However, the major challenge to scale down Cherenkov detectors arises from the large footprint of photon detection apparatus.

In recent decades, surface waves have drawn extensive attention in nanotechnology owing to their ability to confine electromagnetic waves in subwavelength domains [16-19]. Surface plasmon-polariton (SPP) existing at metal/dielectric interfaces is perhaps the most well-known example [20-23]. Its enhanced photonic density of states leads to a wealth of nanotechnological breakthroughs in super-resolution imaging, ultrasensitive biosensing, efficient light emission, etc [24-26]. Especially, recent advances in particle physics have shown that surface plasmon-polaritons may also provide a route to control Cherenkov radiation at the subwavelength scale. This so-called surface-polariton Cherenkov radiation exhibits several unique features including enhanced photon emission, reversed Cherenkov cone and vanishing Cherenkov threshold [12,13,27-35]. However, realization of miniaturized Cherenkov detectors with surface plasmons is still challenging, probably because of the significant dissipation losses and strong chromatic dispersion of metals [36]. For instance, the large metallic dissipation inevitably prohibits the detection of surface-polariton Cherenkov radiation beyond the plasmon propagation length (which is usually several micrometers at optical frequencies). On the other hand, the strong chromatic dispersion makes the emission angle of surface-polariton Cherenkov radiation highly wavelength-dependent and therefore unavoidably limits the working bandwidth of designed Cherenkov detectors. These problems become more severe when the particle velocity approaches the Cherenkov threshold, where the photon emission is inherently weak. Hence, an alternative approach with suppressed dissipation and dispersion effects is still highly demanded for the design of compact Cherenkov detectors.

Here, we show that Dyakonov surface waves (DSWs) supported by transparent birefringent materials provide a feasible solution to these problems[37]. We investigate systematically surface Dyakonov-Cherenkov radiation at the surface of a birefringent crystal in contact with an isotropic background medium and find that the emission behaviors are strongly susceptible to the particle velocity and trajectory. Remarkably, the excitation of DSWs significantly enhances the photon emission by several orders of magnitude when the particle velocity approaches to the threshold velocity of the swift charge in surrounding medium. This unique feature greatly facilitates the particle identification around the velocity cutoff. In addition, DSWs have another two distinct advantages over conventional surface plasmons: first, their negligible dissipation losses not only enhance the photon extraction efficiency, but also significantly increase the propagation length of surface waves, facilitating the far-field detection of Cherenkov signals; second, their small chromatic dispersion dramatically broadens the working bandwidth of particle detectors. Our studies add a new perspective on enhanced particle-photon-matter interactions and open up an opportunity for achieving high-performance Cherenkov detectors on chip.

Without loss of generality, we consider a swift charged particle of velocity $\bar{v}$ travelling parallel to the interface of an isotropic background medium (with a refractive index $n_a$) and a uniaxial birefringent crystal (with ordinary and extraordinary refractive indices denoted as $n_o$ and $n_e$, respectively). The distance between the particle trajectory and the interface is $y_0 = 200$ nm. The optical axis of birefringent crystal is orientated parallel to the interface, as shown in Fig. 1(a). In this configuration, DSWs exist if $n_e > n_a > n_o$ [38-42]. To satisfy this condition, we choose $Si_3N_4$ as the isotropic background medium and $YVO_4$ as the birefringent crystal. In the wavelength range $\lambda \in [0.2, 2]$ µm of our interest, the corresponding refractive indices are given by $n_a = 2.05$, $n_o = 1.99$ and $n_e = 2.22$. In general, DSWs are highly directional and exist only within small angular regions in four quadrants (i.e., $\theta_d$, $\pi - \theta_d$, $\pi + \theta_d$, $2\pi - \theta_d$), where $\theta_d \in [29.43°, 30.15°]$ is the angle between the phase velocity of DSWs and the optical axis of the YVO$_4$ crystal. In this study, we focus our discussions on the behaviors of DSWs in the first quadrant. Owing to the mirror symmetry, our results also apply to DSWs in other three quadrants.

We begin our analysis by studying the excitation condition of DSWs by the swift charged particle. Denote $\theta_q$ as the angle between the optical axis and the particle trajectory [Fig. 1(a)]. We find that the excitation of Dyakonov surface modes requires $\theta_q$ fulfilling the following condition [Supplementary Section S4]:

$$\theta_q = \theta_d \pm \cos^{-1}\left(\frac{\omega}{k_d v}\right), \qquad (1)$$

where $\omega$ is the angular frequency; $k_d$ is the magnitude of in-plane wavevector of the Dyakonov surface mode. Unlike conventional Cherenkov photons which can be produced by charged particles travelling along any direction (i.e. regardless of $\theta_q$), Dyakonov surface modes can be excited only for some specific $\theta_q$.

The radiation field pattern from the swift charged particle is very susceptible to the magnitude and direction of the particle velocity in our platform. To illustrate this point and reveal the impact of DSWs, we plot in Figs. 1(b)-(d) radiation field patterns for three circumstances. When Eq. (1) is rigorously satisfied, the radiation mode is a superposition of the conventional Cherenkov photons and DSWs, as shown in Fig. 1(b). In this case, the radiation field can penetrate deeply into the YVO$_4$ crystal. Such a large penetration length results from the weak longitudinal confinement of DSWs on the interface [17]. In addition, surface Dyakonov-Cherenkov radiation also features an extremely asymmetric field pattern in the transverse plane [Fig. S5]. On the other hand, when the magnitude or direction of the particle velocity changes slightly such that Eq. (1) is no longer satisfied, the swift charged particle emits only conventional Cherenkov photons. As a result, the radiation field decays rapidly in the YVO$_4$ crystal [Figs. 1(c)-(d)] and the field pattern becomes symmetric in the transverse plane [Fig. S5]. These results clearly demonstrate that the excitation of DSWs relies heavily on the particle trajectory. This property provides an effective way to determine simultaneously the magnitude and direction of the particle velocity through the direct measurement of the radiation field pattern.

Excitation of DSWs modifies not only the near-field pattern, but also the energy loss of the swift particle. As shown in Fig. 2, the total radiation power is quite susceptible to the particle velocity and trajectory, and increases dramatically when Eq. (1) is satisfied. To explore the underlying physical mechanism, we divide the total radiation power into two parts, i.e., the radiation loss $P_{\text{ph}}$ through the emission of free-space Cherenkov photons and the radiation loss $P_{\text{Dya}}$ through the excitation of DSWs.

Figures. 2(b)-(c) clarify quantitatively the respective contributions of $P_{\text{ph}}$ and $P_{\text{Dya}}$ to the total radiation power. Our results show that $P_{\text{ph}}$ and $P_{\text{Dya}}$ display distinctly different responses to the particle velocity and trajectory. On the one hand, $P_{\text{ph}}$ (as denoted as the straight dashed line) increases smoothly as the particle velocity increases [Fig. 2(b)] while at the same time remains almost a constant over a broad angular band [Fig. 2(c)]. Such a behavior makes particle detection with traditional Cherenkov photons difficult. On the other hand, $P_{\text{Dya}}$ is much more sensitive to small variations in particle velocity/trajectory and acquires a nonzero value only when $v$ and $\theta_q$ strictly satisfy the condition given by Eq. (1) [see Fig. 2(a) and the bulges in Figs. 2(b)-(c)]. The velocity range for nonzero $P_{\text{Dya}}$ is generally smaller than $0.02c$, e.g., from $0.502c$ to $0.498c$, $0.596c$ to $0.604c$, $0.694c$ to $0.706c$, and $0.792c$ to $0.808c$ for $\theta_q = 41.2°, 65.0°, 75.3°$, and $82.0°$, respectively; the angular band for nonzero $P_{\text{Dya}}$ is generally smaller than $1°$, e.g., from $40.7°$ to $41.7°$, $64.5°$ to $65.5°$, $74.8°$ to $75.8°$, and $81.5°$ to $82.5°$ for $v = 0.5c, 0.6c, 0.7c$, and $0.8c$, respectively. The enhanced

sensitivity in energy loss offers an alternative approach to measure simultaneously the magnitude and direction of the particle velocity.

The negligible chromatic dispersion and small dissipation loss of our structure can greatly enhance the photon extraction by means of DSWs, facilitating particle detection in the far field. To highlight this point, we compare the power flow density of surface Dyakonov-Cherenkov radiation with that of surface-polariton Cherenkov radiation [43]. In our comparison, the surface-polariton Cherenkov radiation is investiageted in the plasmonic system made of Aluminum Oxide ($Al_2O_3$) and Silver (Ag), such that the corresponding SPPs have a propagation constant identical to that of DSWs at a wavelength of $\lambda = 500$ nm [Fig. S8]. Here, the dielectric constants of $Al_2O_3$ and Ag are taken from the experimental data [44]. Figure 3 demonstrates that DSWs are more strongly excited than conventional SPPs despite of their weak confinement along the longitudinal direction. First, the photon extraction efficiency of DSWs is much higher (i.e. the power flow density is several orders of magnitude larger) than that of conventional SPPs at $l = 0$ μm, as shown in Fig. 3(c). Second, DSWs have a much longer propagation distance and remain detectable in the far field, e.g. the Poynting power of DSWs attenuates less than 7% over a distance of 20 μm, while conventional SPPs have already faded away over such a distance.

We demonstrate that DSWs offer a powerful platform to manipulate Cherenkov radiation at the nanoscale. Different from conventional Cherenkov or surface polaritons-Cherenkov radiation, surface Dyakonov-Cherenkov radiation is highly susceptible to the particle trajectory. For the practical implementation, the optical axis orientation (and hence $\theta_q$) can be effectively tuned by a Galvo motor [45], providing a route to control surface Dyakonov-Cherenkov radiation. Small chromatic dispersion and negligible propagation loss make our all-dielectric structure highly suitable for particle detection on chip.

**Acknowledgements**

Y.L. was sponsored in part by Singapore Ministry of Education (No. MOE2018-T2-2-189 (S), MOE2017-T1-001-239 (RG91/17 (S)), A*Star AME Programmatic Funds (No. A18A7b0058) and National Research Foundation Singapore Competitive Research Program (No. NRF-CRP18-2017-02). B.Z. was sponsored in part by Singapore Ministry of Education (No. MOE2018-T2-1-022 (S), MOE2016-T3-1-006 and Tier 1 RG174/16 (S)). L.J.W. was sponsored in part by the Advanced Manufacturing and Engineering Young Individual Research Grant (No. A1984c0043) from the Science and Engineering Research Council of the Agency for Science, Technology and Research, Singapore.

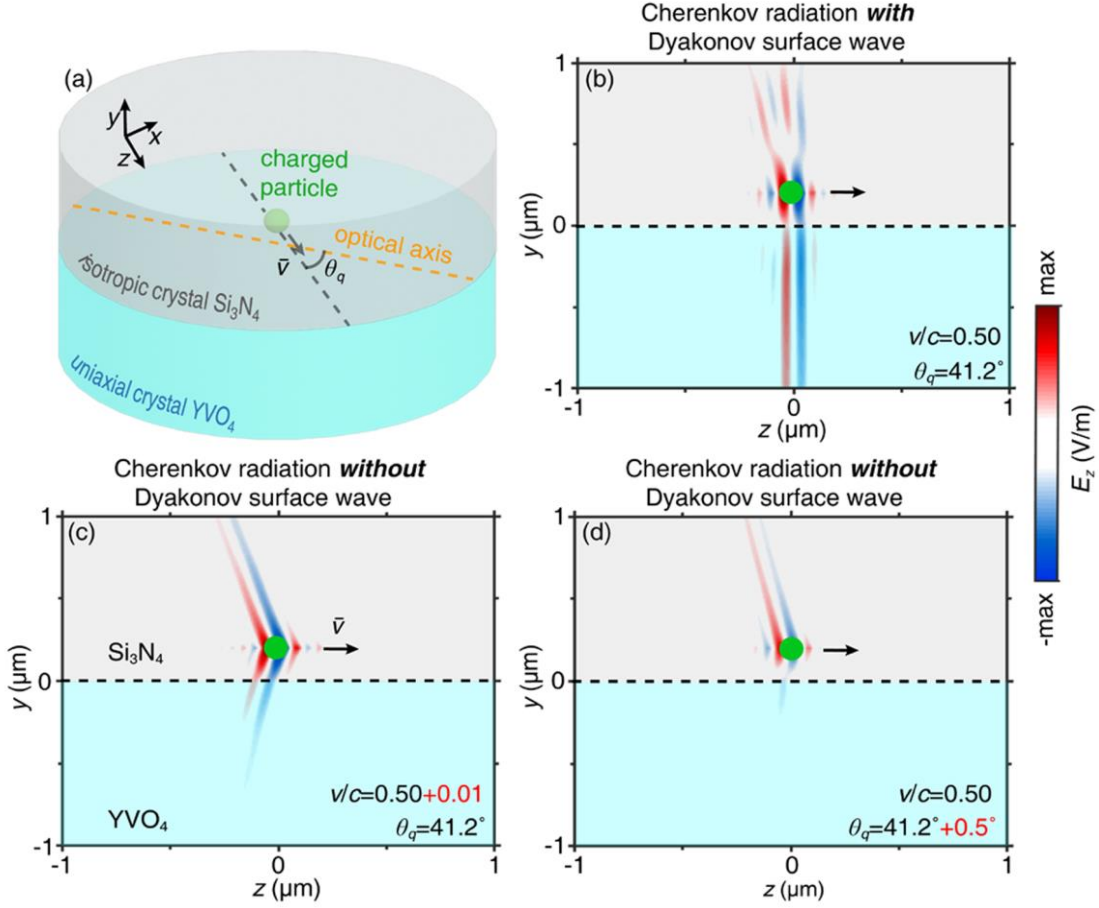

FIG. 1. Emission behaviors of a swift charged particle in an all-dielectric structure made of a semi-infinite istrotropic medium $Si_3N_4$ and a semi-infinite uniaxial crystal $YVO_4$. (a) The schematic of the setup. A swift charged particle moves parrallel to the interface between the isotropic medium $Si_3N_4$ and the uniaxial crystal $YVO_4$. The angle between the particle velocity and the optical axis of the uniaxial crystal is denoted as $\theta_q$. (b-d) Time-domain radiation-field distributions of the charged particle in the cross section formed by the surface normal and the particle trajectory. When $v = 0.50c$ and $\theta_q = 41.2°$, the swift charged particle excites both free-space Cherenkov photons and DSWs in (b); When $v$ is changed to $0.51c$ [in (c)] or $\theta_q$ is changed to be $41.7°$ [in (d)], only free-space Cherenkov photons are produced. In all the plots, the distance between the particle trajectory and the interface is fixed as $y_0 = 200$ nm, and the integration range of the wavelength is $\lambda \epsilon [0.2,2]$ μm in free space.

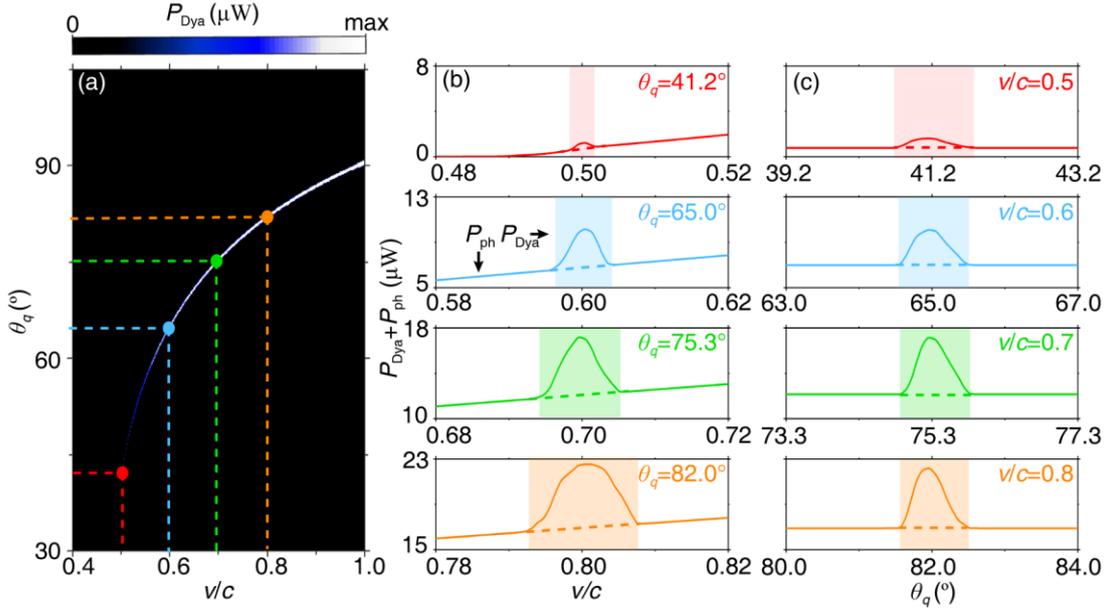

FIG. 2. Energy loss of a swift charged particle emitting Dyakonov surface wave. (a) The radiation loss $P_{\text{Dya}}$ (owing to the excitation of DSWs) as a function of the normalized particle velocity $v/c$ and the angle $\theta_q$ (corresponding to the orientation of the particle trajectory). $P_{\text{Dya}}$ acquires a nonzero value only when $v$ and $\theta_q$ satisfy the phase matching condition given by Eq. (1). (b) The total energy loss $P_{\text{Dya}} + P_{\text{ph}}$ (where $P_{\text{ph}}$ refers to the radiation loss of free-space Cherenkov photons) versus the normalized particle velocity $v/c$ for angle $\theta_q = 41.2°$, $65.0°$, $75.3°$ and $82.0°$, respectively. (c) The total energy loss $P_{\text{Dya}} + P_{\text{ph}}$ versus the angle $\theta_q$ for particle velocity $v = 0.5c$, $0.6c$, $0.7c$ and $0.8c$, respectively. In particular, we denote $P_{\text{ph}}$ as the straight dashed line (b-c).

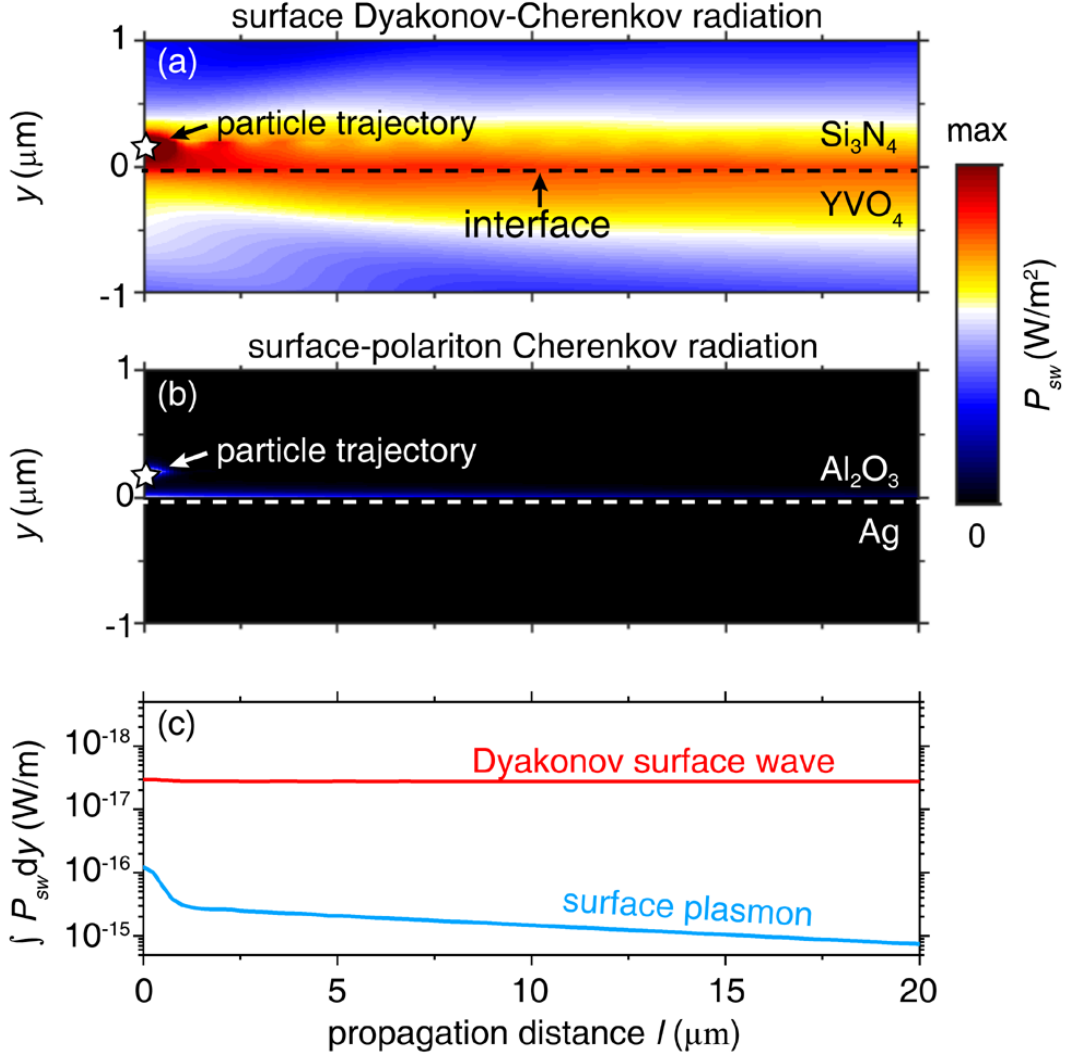

FIG. 3. Comparison between surface Dyakonov-Cherenkov radiation and surface-polariton Cherenkov radiation. (a) The contour plot of the power flow density in the cross section formed by the surface normal and the group velocity direction of DSWs. (b) The contour plot of the power flow density in the cross section formed by the surface normal and the propagation direction of conventional SPPs. (c) The Poynting power $\bar{P}_{sw}$ integrated along the $y$-direction versus the propagation distance $l$. In our comparison, we choose $Si_3N_4$ - $YVO_4$ and $Al_2O_3$ - Ag as the platforms to excite DSWs and SPPs, respectively, such that the corresponding DSWs and SPPs have an identical effective mode index $n_{\text{eff}} = 2.0403$ at the wavelength of $\lambda = 500$ nm. The particle velocity is $v = 0.5c$ for both configurations. As a result, DSWs/SPPs propagate in a direction with 14.21°/11.40° to the particle trajectory.